\documentclass[prc,twocolumn,tightenlines
,amssymb,aps,nobibnotes,superscriptaddress,showpacs,balancelastpage,floatfix]{revtex4-1}
\usepackage{graphicx}

\begin{document}
\title{Lifetime measurements in $^{63}$Co and $^{65}$Co}
\author{A.~Dijon}
\email{dijon@ganil.fr}
\affiliation {Grand Acc\'el\'erateur National d'Ions Lourds (GANIL), CEA/DSM-CNRS/IN2P3, Boulevard H. Becquerel, F-14076, Caen, France}
\author{E.~Cl\'ement}
\affiliation {Grand Acc\'el\'erateur National d'Ions Lourds (GANIL), CEA/DSM-CNRS/IN2P3, Boulevard H. Becquerel, F-14076, Caen, France}
\author{G.~de~France}
\affiliation {Grand Acc\'el\'erateur National d'Ions Lourds (GANIL), CEA/DSM-CNRS/IN2P3, Boulevard H. Becquerel, F-14076, Caen, France}
\author{P.~Van~Isacker}
\affiliation {Grand Acc\'el\'erateur National d'Ions Lourds (GANIL), CEA/DSM-CNRS/IN2P3, Boulevard H. Becquerel, F-14076, Caen, France}
\author{J.~Ljungvall}
\affiliation {Grand Acc\'el\'erateur National d'Ions Lourds (GANIL), CEA/DSM-CNRS/IN2P3, Boulevard H. Becquerel, F-14076, Caen, France}
\affiliation {CSNSM, CNRS/IN2P3, 91400 Orsay, Cedex}
\affiliation {CEA Saclay, IRFU, SPHN, F-91191 Gif-sur-Yvette, France}
\author{A.~G\"orgen}
\affiliation {CEA Saclay, IRFU, SPHN, F-91191 Gif-sur-Yvette, France}
\affiliation {Department of Physics, University of Oslo, N-0316 Oslo, Norway}
\author{A.~Obertelli}
\affiliation {CEA Saclay, IRFU, SPHN, F-91191 Gif-sur-Yvette, France}
\author{W.~Korten}
\affiliation {CEA Saclay, IRFU, SPHN, F-91191 Gif-sur-Yvette, France}
\author{A.~Dewald}
\affiliation {Institut f\"ur Kernphysik, Universit\"at zu K\"oln, D-50937 K\"oln, Germany}
\author{A.~Gadea}
\affiliation {Instituto de F\'isica Corpuscular, CSIC-University of Valencia, E-46071 Valencia, Spain}
\author{L.~Gaudefroy}
\affiliation {CEA, DAM, DIF, F-91297 Arpajon, France}
\author{M.~Hackstein}
\affiliation {Institut f\"ur Kernphysik, Universit\"at zu K\"oln, D-50937 K\"oln, Germany}
\author{D.~Mengoni}
\affiliation {Dipartimentito di Fisica dell'Universit\`a and INFN, I-35131 Padova, Italy}
\affiliation {University of the West of Scotland, Paisley, UK}
\author{Th.~Pissulla}
\affiliation {Institut f\"ur Kernphysik, Universit\"at zu K\"oln, D-50937 K\"oln, Germany}
\author{F.~Recchia}
\affiliation {Dipartimentito di Fisica dell'Universit\`a and INFN, I-35131 Padova, Italy}
\author{M.~Rejmund}
\affiliation {Grand Acc\'el\'erateur National d'Ions Lourds (GANIL), CEA/DSM-CNRS/IN2P3, Boulevard H. Becquerel, F-14076, Caen, France}
\author{W.~Rother}
\affiliation {Institut f\"ur Kernphysik, Universit\"at zu K\"oln, D-50937 K\"oln, Germany}
\author{E.~Sahin}
\affiliation {LNL(INFN), Laboratori Nazionali di Legnaro, I-35020 Legnaro, Italy}
\author{C.~Schmitt}
\affiliation {Grand Acc\'el\'erateur National d'Ions Lourds (GANIL), CEA/DSM-CNRS/IN2P3, Boulevard H. Becquerel, F-14076, Caen, France}
\author{A.~Shrivastava}
\affiliation {BARC, Nuclear Physics Division, Trombay, Mumbai 400085, India}
\author{J.J.~Valiente-Dob\'on}
\affiliation {LNL(INFN), Laboratori Nazionali di Legnaro, I-35020 Legnaro, Italy}
\author{K.O.~Zell}
\affiliation {Institut f\"ur Kernphysik, Universit\"at zu K\"oln, D-50937 K\"oln, Germany}
\author{M.~Zieli\'nska}
\affiliation {Heavy Ion Laboratory, Warsaw University, Warsaw, PL-02097, Poland}

\date{\today}

\begin{abstract}
Lifetimes of the $9/2^-_1$ and $3/2^-_1$ states in $^{63}$Co
and the $9/2^-_1$ state in $^{65}$Co
were measured using the recoil distance Doppler shift
and the differential decay curve methods.
The nuclei were populated by multi-nucleon transfer reactions in inverse kinematics.
Gamma rays were measured with the EXOGAM Ge array
and the recoiling fragments were fully identified
using the large-acceptance VAMOS spectrometer.
The E2 transition probabilities
from the $3/2^-_1$ and $9/2^-_1$ states to the $7/2^-$ ground state
could be extracted in $^{63}$Co
as well as an upper limit for the $9/2^-_1\rightarrow7/2^-_1$ $B$(E2) value in $^{65}$Co.
The experimental results were compared to large-scale shell-model calculations
in the $pf$ and $pfg_{9/2}$ model spaces,
allowing to draw conclusions on the single-particle or collective nature of the various states.
\end{abstract}


\maketitle
\section{Introduction}
\label{s_intro}
Doubly-magic nuclei are the cornerstones of the nuclear shell model.
In $^{68}$Ni, the observation of a first-excited $0^+$ state at low energy~\cite{bernas_magic_1982}, 
the high excitation energy of the $2^+_1$ state
and the low $B({\rm E2};2^+_1\rightarrow0^+_1)$ value
are evidence for the doubly-magic character
of this nucleus~\cite{broda_n40_1995,sorlin__28^68ni_40:_2002,bree_coulomb_2008}.  
Consequently, $^{68}$Ni and its immediate neighbors
have attracted the interest of many experimental and theoretical groups.

It has been shown experimentally that the magicity at $N=40$ is localized
and disappears as soon as a few nucleons
are added to or removed from the $^{68}$Ni core.
Coulomb-excitation experiments using a $^{70}$Ni radioactive beam
have shown an important increase in collectivity
when neutrons start filling the $1\nu g_{9/2}$ orbital.
This has been interpreted as a `polarization' of the $^{68}$Ni core~\cite{perru_enhanced_2006}
which results from a reduced splitting
between the $1\pi f_{7/2}$ and $1\pi f_{5/2}$ proton orbitals
in the Ni isotopes beyond $N=40$,
leading to a weakening of the $Z=28$ subshell closure.
This interpretation has been further reinforced
by recent mass measurements~\cite{gunaut_high-precision_2007}.

Above the Ni chain, when protons are added to $^{68}$Ni,
the spectroscopy of $^{69,71,73}$Cu
has revealed states with a rich variety of characters~\cite{stefanescu_interplay_2008} 
and a delicate interplay between single-particle and collective structures.
The observed spectra for all three cupper isotopes include
a $5/2^-$ state with single-particle character,
a collective $1/2^-$ state
and a $7/2^-$ state resulting
from the coupling of a single proton in the $2\pi p_{3/2}$ orbital
to the $2^+_1$ state in the Ni core.
It seems therefore that a single Cu isotope
contains coexisting single-particle and collective states.
Below the Ni chain, the spectroscopy of Fe isotopes
has shown an increased collectivity toward $N=40$,
revealing the collapse of this subshell closure~\cite{hannawald_decay_1999}.
Very recently, the $2^+_1\rightarrow0^+_1$ transition probabilities in $^{62,64,66}$Fe
have been determined from the lifetime of the $2^+_1$ state~\cite{ljungvall_onset_2010,rother_enhanced_2011}.
The increase in the $B({\rm E2};2^+_1\rightarrow0^+_1)$ value at $N=38$
confirms this onset of collectivity
which can be explained in the shell model
only if both the $1\nu g_{9/2}$ and the $2\nu d_{5/2}$ orbitals are included.
By analogy, since Cu and Co are mirror chains around semi-magic Ni,
one therefore also expects the occurrence
of coexisting single-particle and collective states in Co.

Recent experimental results have provided new information
on the structure of neutron-rich Co isotopes~\cite{pauwels_structure_2009}.
It was established that the $9/2^-_1$ level in Co and the $2^+_1$ level in Ni
follow the same energy systematics with neutron number.
Hence, the $9/2^-_1$ state conceivably can be considered
as a proton hole $(1\pi f_{7/2})^{-1}$ coupled to $2^+_1({\rm Ni})$.
In contrast, the energy systematics of the $3/2^-_1$ level
in odd-mass neutron-rich Co nuclei
mimics the behavior of the $2^+_1$ level in the Fe isotopes.
Consequently, the $3/2^-_1$ state might be interpreted
as a single $1\pi f_{7/2}$ proton coupled to $2^+_1({\rm Fe})$.
No such simple conclusions can be drawn
on the basis of known E2 transition probabilities.

The purpose of the present study is to investigate if information on E2 and M1 transitions,
as can be derived from lifetime measurements,
can shed light on the character of states in the Co isotopes
and might meaningfully constrain nuclear models in this mass region.
We report on
lifetime measurements of the $9/2^-_1$ and $3/2^-_1$ states in $^{63}$Co
and of the $9/2^-_1$ state in $^{65}$Co,
using the recoil distance Doppler shift technique~\cite{dewald_differential_1989}.
This information is used to determine electromagnetic transition probabilities
which in turn elucidate the structure of states in $^{63,65}$Co.

The paper is organized as follows.
The experiment is described in Sect.~\ref{s_expt} 
and the results are presented in Sect.~\ref{s_res}.
The discussion and a comparison with large-scale shell-model calculations 
are presented in Sect.~\ref{s_calc}.
Finally, the conclusions of this work are summarized in Sect.~\ref{s_conc}.

\section{Experimental method}
\label{s_expt}
\begin{figure}[!h]
\centering
\includegraphics[width=8cm,height=6cm]{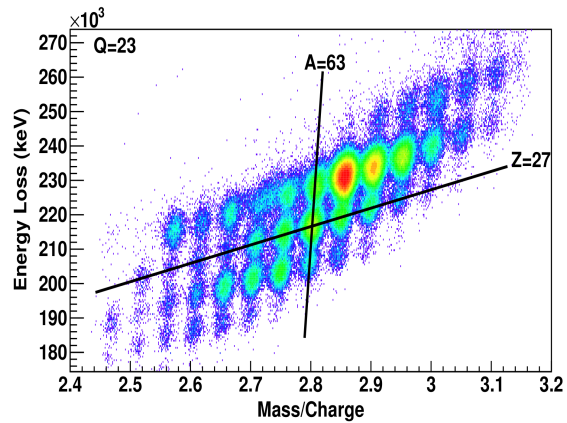}
\caption {(Color online)Identification matrix showing the energy loss in the ionization chamber of VAMOS as a function of the mass-over-charge ratio determined from the time of flight and magnetic rigidity of the spectrometer. This plot is obtained from a single silicon detector and a single charge state Q=23. The crossing of the two lines indicates the position of $^{63}$Co.}
\label{f_vamos}
\end{figure}

The experiment was performed at the Grand Acc\'el\'erateur National d'Ions Lourds (GANIL) 
by applying the recoil distance Doppler shift (RDDS) method~\cite{dewald_differential_1989}
combined with multi-nucleon transfer reactions in inverse kinematics.
A $^{238}$U beam at 6.5 A.MeV
and  a 1.5 mg/cm$^2$ thick $^{64}$Ni target
followed by a thick 4.7~mg/cm$^2$ Mg degrader were used. 
The target-like reaction products were detected and identified
in the large-acceptance VAMOS spectrometer~\cite{pullanhiotan_performance_2008}.
The optical axis of the spectrometer was positioned at $45^\circ$ with respect to the beam axis, such that the grazing angle was within the angular acceptance of the spectrometer.
The focal-plane detection system of VAMOS allows the full reconstruction of the trajectories and an unambiguous identification of the reaction products
in mass, charge and atomic number (see Fig.\ref{f_vamos}).

A dedicated differential plunger device for multi-nucleon transfer reactions has been designed and built at the University of Cologne. In multi-nucleon transfer and deep-inelastic reactions, the recoiling nuclei of interest leave the target in a direction different from the one of the incoming beam. This plunger device can be oriented in such a way that the degrader foil moves along the axis of the VAMOS spectrometer.

\begin{figure}
\centering
\includegraphics[width=9cm,height=6cm]{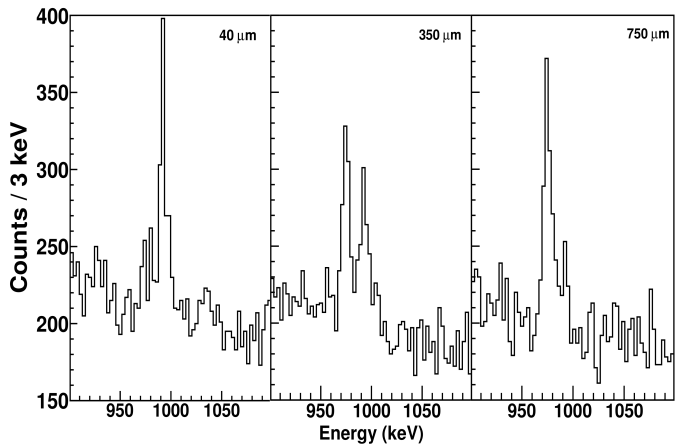}
\caption{
Gamma-ray spectra obtained for the distances $x=40$, 350 and 750~$\mu$m.
The low (high) energy peak corresponds to the fast (slow)component of the $3/2^-_1\rightarrow 7/2^-_1$ transition in $^{63}$Co.}
\label{f_im1}
\end{figure}

Doppler-corrected prompt gamma-rays
emitted before and after the degrader foil
were measured by the high-efficiency hyper pure EXOGAM Ge array~\cite{simpson_exogam_2000,shepherd_measurementsprototype_1999} 
in coincidence with the recoils identified 
in the VAMOS spectrometer.
The EXOGAM array consisted of one detector at $180^\circ$ with respect to the optical axis of VAMOS,
three detectors at $135^\circ$
and five detectors at $90^\circ$.
The latter detectors could not be used to measure Doppler shifts
but were meant to increase the number of $\gamma$--$\gamma$ coincidences 
to lift possible degeneracies arising from transitions with similar energies. 
Data were collected for six target-to-degrader distances: 40, 120, 250, 350, 450 and 750~$\mu$m,
optimized for measuring lifetimes between 1 and 10~ps.
The recoil velocity of the reaction products after the target
was determined with the VAMOS spectrometer by removing the degrader from the plunger. The typical average 
recoil velocity of the target-like products before and after the degrader foil was 35~$\mu$m/ps and 30~$\mu$m/ps 
respectively.  
In the application of the RDDS method,
the following decay curves were built:
\begin{equation}
Q_{ij}(x)=\frac{I^{\rm s}_{ij}(x)}{I^{\rm s}_{ij}(x)+I^{\rm f}_{ij}(x)},
\end{equation}
where $i$ and $j$ are associated with the initial and final states, respectively,
and $I^{\rm s}_{ij}(x)$ and $I^{\rm f}_{ij}(x)$ denote the intensities
of the slow (s) and fast (f) components
occurring between the states $i$ and $j$
for a target-to-degrader distance $x$~\cite{dewald_differential_1989}.
Figure~\ref{f_im1} shows the gamma-ray spectra
containing the $3/2^-_1\rightarrow7/2^-_1$ transition in $^{63}$Co
for three values of $x$.
The evolution of the intensity with the distance $x$ 
in the fast and slow components is clearly visible.
The intensities were determined assuming, for all distances,
the same width and position for the fast (slow) component.
Due to the Doppler correction the two components
are well separated in detectors located at angles larger than $135^\circ$. 

For each distance, the normalization factors $I^{\rm s}_{ij}(x)+I^{\rm f}_{ij}(x)$
were found to be consistent with the number of ions identified in VAMOS.
The lifetime was extracted from decay curves for each distance~\cite{dewald_differential_1989} as
\begin{equation}
\tau_{i}(x)=-\left(v\frac{dQ_{ij}(x)}{dx}\right)^{-1} Q_{ij}(x),
\label{e_decay2}
\end{equation}
with $v$ the recoil velocity. As explained in detail later, the analysis was performed in such a way that 
the feeding of the state of interest from higher-lying states is suppressed.
Differentiable polynomials were fitted to the decay curve
in order to determine their derivatives.
The differential decay curve method
gives an independent value for the lifetime
for each target-to-degrader distance that lies within the sensitivity range
for this particular lifetime according to eq.~(\ref{e_decay2}). 

\section{Results}
\label{s_res}

\begin{table}
\caption{
Summary of results for the $9/2^-_1$ state in $^{63,65}$Co
and the $3/2^-_1$ state in $^{63}$Co.}
\label{t1} 
\begin{ruledtabular}
\begin{tabular}{ccrrc}
\noalign{\smallskip}
& $J^\pi$ & $E_{\rm exp}$~(keV) & $\tau_{\rm exp}$~(ps) & $B({\rm E2},\downarrow)_{\rm exp}$(W.u)\\
\noalign{\smallskip}\hline\noalign{\smallskip}
$^{63}$Co & $3/2^-_1$ &   995.1 & 15.4~(18)  & 3.71~(43)\phantom{$^a$}\\
                  & $9/2^-_1$ & 1383.5 & 0.9~(4)    & 12.2~(54)$^a$\\
$^{65}$Co & $9/2^-_1$ & 1479.4 & $\leq17.3$ & $\geq0.43^a$\\
\noalign{\smallskip}
\multicolumn{5}{l}{$^a$Assuming a pure E2 transition.}
\end{tabular}
\end{ruledtabular}
\end{table}
A summary of the results of our experiment is shown in table~\ref{t1}.
We now discuss these results separately for the two isotopes,
$^{63}$Co and $^{65}$Co.
\begin{figure}
\centering
\includegraphics[width=9cm,height=6cm]{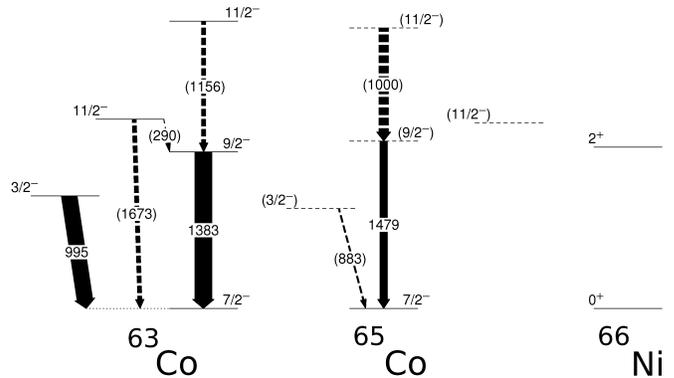}
\caption{
Levels in $^{63}$Co and $^{65}$Co
connected by transitions identified in this experiment.
Full arrows indicate observed transitions from which the lifetime was extracted
and dashed arrows indicate observed transitions from which it was not possible to extract the lifetime.
The $2^+_1$ excitation energy in $^{66}$Ni is shown for comparison.}
\label{f_level}
\end{figure}

In $^{63}$Co, it was possible to identify five transitions:
$3/2^-_1\rightarrow7/2^-_1$,
$9/2^-_1\rightarrow7/2^-_1$,
$11/2^-_1\rightarrow7/2^-_1$, $11/2^-_2\rightarrow9/2^-_1$, and $11/2^-_1\rightarrow9/2^-_1$.
Only for the first two transitions there was enough statistics to deduce a lifetime.
From the $3/2^-_1\rightarrow7/2^-_1$ transition,
the lifetime of the $3/2^-_1$ state was extracted assuming direct feeding (see Fig.~\ref{f_level}).
A low $B$(E2) value is obtained from this lifetime (see table 1)
which is indicative of the weak collective character of the $3/2^-_1$ state.
For the $9/2^-_1\rightarrow7/2^-_1$ transition, whatever the target-to-degrader distance was and without any 
selection of the events, the slow component was found systematically more intense than the fast one. 
Consequently, the extracted lifetime was found surprisingly long with a value of 35 ps.

The most likely origin of such a long apparent lifetime 
is the presence of a long-lived state 
in the feeding path of the $9/2^-_1$ level. 
Unfortunately, the number of $\gamma$--$\gamma$ coincidences
was not large enough to establish the feeding path.
Alternatively, we have exploited the capability of the VAMOS spectrometer 
to reconstruct the full kinematics of the recoils on an event-by-event basis.  
In particular, the excitation energy of the recoils can be estimated from 
selection gates in the two-dimensional plot of the total kinetic energy 
as a function of the recoil scattering angle ~\cite{rejmund_shell_2007}. 

\begin{figure*}[!ht]
\centering
\includegraphics[width=16cm,height=10cm]{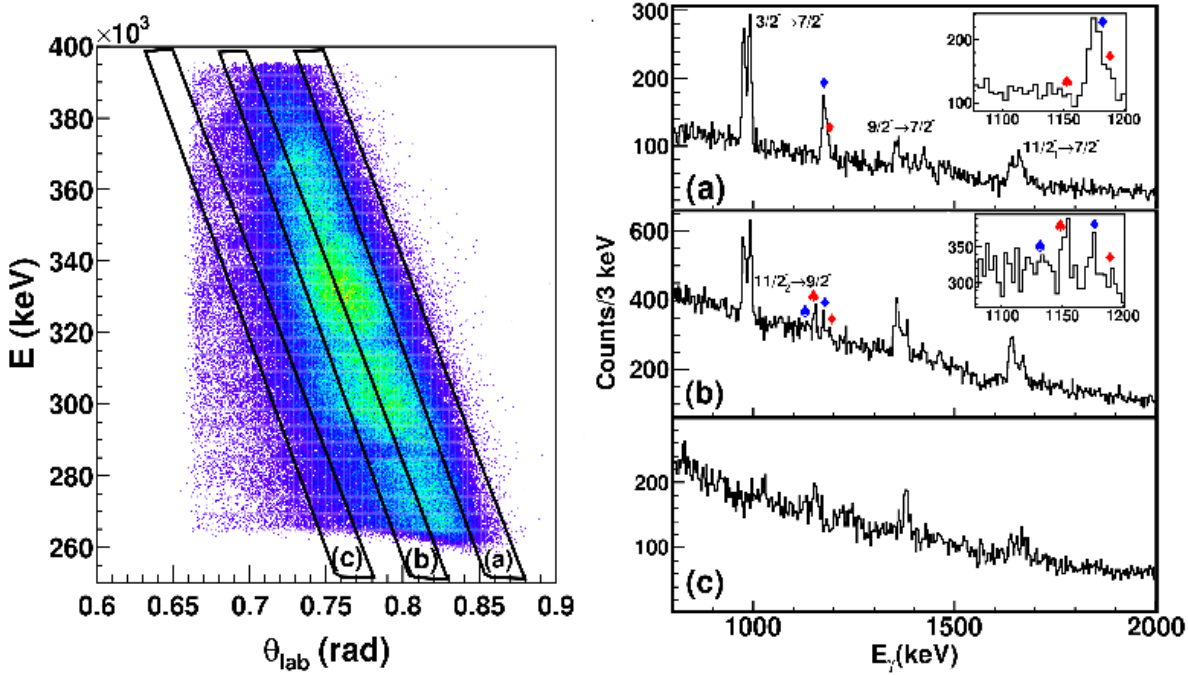}
\caption{
(Color online) Left: Total kinetic energy as a function of the scattering angle in the 
laboratory frame. The slices superimposed on the 2D plot show examples of cuts 
in excitation energy. Right: Gamma-ray spectra obtained for detectors located at angles larger than $115^\circ$, for all distances and gated by the excitation energy gates indicated on the left. Gate (a) corresponds to recoils with the lowest energy while gate (c) selects events with the largest one. One clearly sees the evolution of the spectra while going from (a) to (c). The inset for gates (a) and (b) shows the region of the $11/2^-_2\rightarrow9/2^-_1$. In the inset of (a), we note the presence of a peak at 1187 keV which is not known in $^{63}$Co.The red square probably corresponds to the slow component as the blue one corresponds to the fast one. The red spade indicates the position of the unobserved slow component of the decay of the 11/2$^{-}_{2}$ state. The strong suppression in (b) of the 1187 keV transition suggests a transition to the ground state. In the inset of (b), the red spade shows the slow component of the 1156 keV transition and the remainder of the 1187 keV fast component. The blue spade indicates the position of the unobserved fast component of the 1156 keV.}
\label{f_im2}
\end{figure*}

This is illustrated in Fig.~\ref{f_im2} which shows such a 
matrix. The three $\gamma$-ray spectra are obtained using the 
three different gates shown on the two-dimensional plot. 
As seen in the gamma-ray spectra of $^{63}$Co, a high (low) excitation 
energy selection, completely removes the decay from the $3/2^-_1$ ($11/2^-_2$) states, respectively. This allows us to control the feeding of the $9/2^-_1$ 
state (see also~\cite{valiente-dobn_lifetime_2009,the_clara-prisma_collaboration_lifetime_2009}). 
We then defined a low excitation energy gate, which preserves the direct feeding of this state 
while suppressing feeding from above, and a high energy gate. Using the 290 keV transition ($11/2^{-}_{1}\rightarrow 9/2^{-}$), we measured that the unobserved
 feeding $11/2^{-}_{2}\rightarrow 9/2^{-}$ is less than $6\%$.  
We estimated that the low energy gate selects events with an 
excitation energy between 0.5 and 2.3 MeV. The boundary between the 
two regions is further confirmed by the convergence of the extracted lifetime to the 
shortest value for the $9/2^-_1$ state (see later).

The various selections in excitation energy
do not influence the lifetime extracted for the $3/2^-_1$ state,
confirming that there is no long-lived level feeding this particular state.
In contrast, the gamma-ray spectra
obtained for the $9/2^-_1\rightarrow7/2^-_1$ transition
using the low or the high excitation energy gate are very different.
For the high excitation energy selection,
most of the decays occur after the degrader,
indicating a rather long lifetime.
By applying the low excitation energy gate, we suppress the feeding of $9/2^-_1$ from  
higher-lying states and 
most of the decays occur before the degrader
corresponding to a short lifetime of 0.9~(4)~ps.
The comparison of the low and high excitation energy selection
suggests the presence of a level with a lifetime
of several hundreds of picoseconds in the feeding path of the $9/2^-_1$ state. 

There are several possible candidates for this unknown long-lived level:
(i) One candidate is the $11/2^-_2$ level which lifetime is unknown.
(ii) Another possibility is that this state decays by the 2050 keV transition
which is observed but cannot be placed in the level scheme of $^{63}$Co.
However, with such a high energy,
the lifetime of the decaying level would be much shorter.
(iii) It can also be a long-lived state at relatively high excitation energy
decaying by a low-energy gamma ray, eventually reaching the $9/2^-_1$ level. 
Indeed, a gamma ray with an energy below 115 keV
would have been fully attenuated by the absorbers,
installed in front of the germanium detectors
and needed to reduce the counting rates in EXOGAM
coming from X-rays
and from gamma transitions between members
of a low-lying rotational band in $^{238}$U,
populated by Coulomb excitation. However, again, this alternative is very unlikely.

Unfortunately, due to the low statistics it was not possible to obtain an E2/M1 ratio from angular distribution for the $9/2^-_1\rightarrow7/2^-_1$ transition
which in turn prevented us to extract a $B$(E2) value. On the basis of the possible core excitation nature of the $9/2^-_1$ state, one may expect a large quadrupole contents for such a transition. Assuming a pure quadrupole nature, a rather collective $B$(E2) value of 12.2~(54)~W.u is obtained for this transition.

In $^{65}$Co, we identified three transitions:
$3/2^-_1\rightarrow7/2^-_1$,
$9/2^-_1\rightarrow7/2^-_1$
and $11/2^-_2\rightarrow9/2^-_1$ but 
only the lifetime for the $9/2^-_1\rightarrow7/2^-_1$ transition could be measured.
As in the previous case, a very long lifetime was extracted for this transition
which showed a dependence on the excitation energy selection,
indicating the presence of a long-lived state in the feeding path.
An upper limit of 17.3~ps was obtained
for the lifetime of the $9/2^-_1$ level in $^{65}$Co (see table~\ref{t1})
corresponding to a lower limit  of $B({\rm E2};9/2^-_1\rightarrow7/2^-_1)\geq0.43$~W.u,
if a pure E2 transition is assumed.

\section{Discussion and interpretation}
\label{s_calc}
\begin{figure*}[!ht]
\centering
\includegraphics[width=16cm,height=10cm]{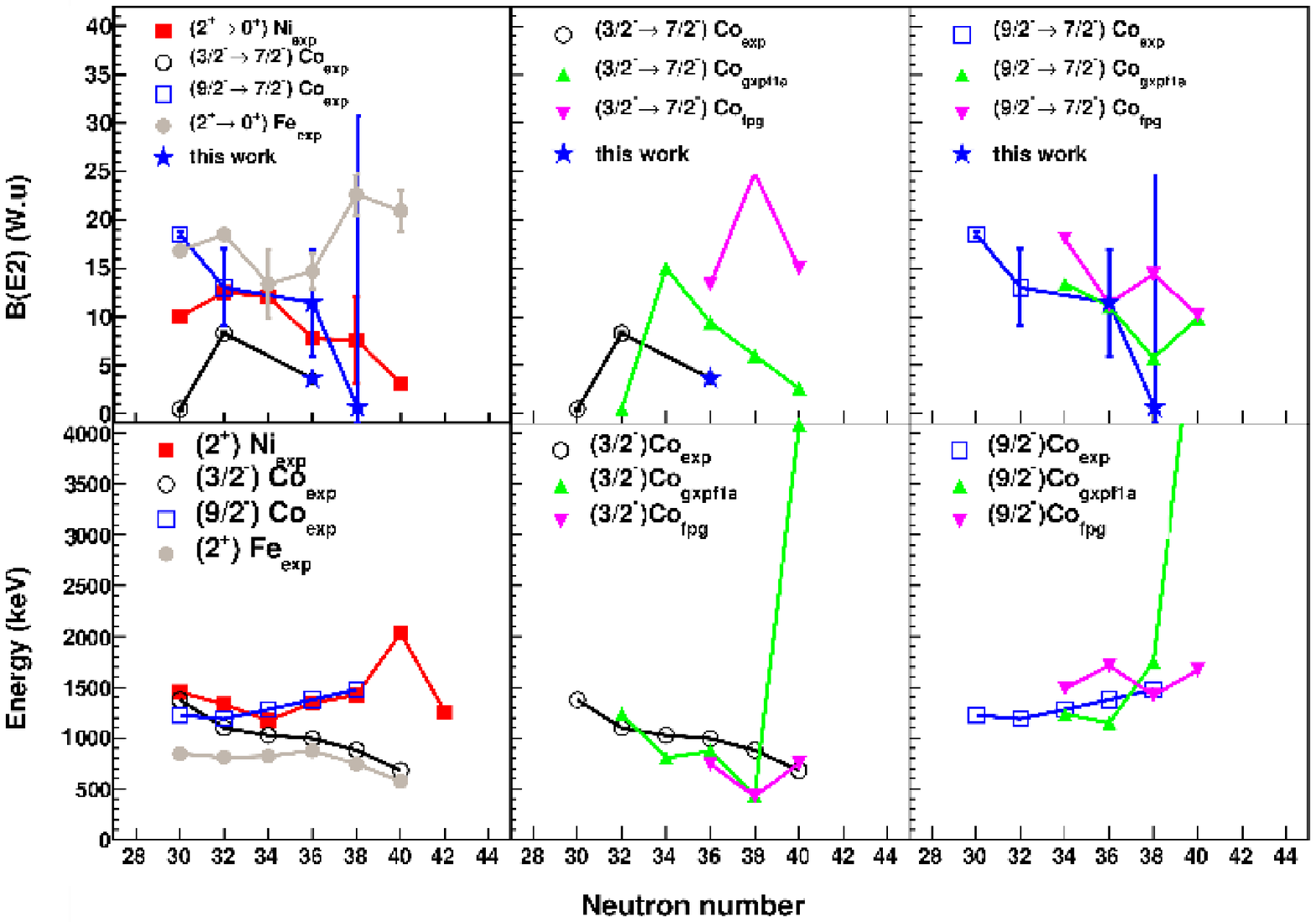}
\caption{
(Color online) Left panels:
Systematics of the $B$(E2) values (a) and the experimental energies (b) in the even-$N$ Fe, Co and Ni isotopes.
Middle and right panels:
Calculated and measured $3/2^-_1$ and $9/2^-_1$ energies (bottom middle and right)
and $B({\rm E2};3/2^-_1\rightarrow7/2^-_1)$
and $B({\rm E2};9/2^-_1\rightarrow7/2^-_1)$ values (top middle and right)
in the odd-mass Co isotopes.}
\label{f_eb}
\end{figure*}

The systematics of excitation energies
and $B$(E2) values in the Fe, Co and Ni isotopes
are shown in Fig.~\ref{f_eb}.
The experimental excitation energies of the $2^+_1$ states
in the even-even Ni and Fe isotopes are shown in lower left panel
together with the energies of the $3/2^-_1$ and $9/2^-_1$ states
in odd-mass Co nuclei.
We observe that the $9/2^-_1$ and $2^+_1({\rm Ni})$ excitation energies
as well as those of the $3/2^-_1$ and $2^+_1({\rm Fe})$ states
behave similarly.
It is therefore tempting to interpret
the $9/2^-_1$ state as a proton hole coupled to $2^+_1({\rm Ni})$~\cite{pauwels_structure_2009}
and the $3/2^-_1$ state as a proton particle coupled to $2^+_1({\rm Fe})$.

The E2 transition probabilities, shown in the upper left panel of Fig.~\ref{f_eb},
are more difficult to explain in this simple picture. 
The $B({\rm E2};3/2^-_1\rightarrow7/2^-_1)$ value in Co
is well below that of $B({\rm E2};2^+_1\rightarrow0^+_1)$ in Fe.
This might indicate that the $N=40$ gap has not completely collapsed at $Z=27$
or that the $3/2^-_1$ state has in fact a single-particle character.
As shown in Ref.~\cite{ljungvall_onset_2010},
the onset of collectivity in Fe is clearly established at $N=38$.
If the $3/2^-_1$ state results from the coupling of a proton to  $2^+_1({\rm Fe})$,
then the $B({\rm E2};3/2^-_1\rightarrow7/2^-_1)$ value should rise at $N=38$,
making the measurement of this quantity in $^{65}$Co important.

The $9/2^-_1\rightarrow7/2^-_1$ transition probability
is even more difficult to interpret
since the decay occurs either by E2 or M1,
and the interpretation depends on the assumed multipolarity.
Assuming a pure E2 transition,
the $B({\rm E2};9/2^-_1\rightarrow7/2^-_1)$ value extracted for $^{63}$Co
is compatible with the $B({\rm E2};2^+_1\rightarrow0^+_1$) value in $^{64}$Ni.

Experimental results were compared
to large-scale shell-model calculations
(carried out with the ANTOINE code~\cite{caurier_shell_2005})
in the odd-mass neutron-rich $^{61-67}$Co isotopes,
using interactions tailored for the $pf$ and $pfg_{9/2}$ valence spaces.
For the former we used the {\sl gxpf1a} interaction of Honma {\it et al.}~\cite{honma_shell-model_2005}
in the space consisting
of the $1f_{7/2}$, $2p_{3/2}$, $1f_{5/2}$ and $2p_{1/2}$ orbitals
with the single-particle energies $-8.624$, $-5.679$, $-1.383$ and $-4.137$~MeV, respectively,
and with $^{40}$Ca as an inert core.
The full model space was considered for $^{61,63,65,67}$Co.
For lighter Co isotopes the model space becomes very large
and no converged results could be obtained.
For the larger valence space the $1g_{9/2}$ orbital was included
and the interaction of Ref.~\cite{sorlin__28^68ni_40:_2002} (which shall be called {\sl pfg}) was used 
with the single-particle energies 0.0, 2.0, 6.5, 4.0 and 9.0~MeV
for the $1f_{7/2}$, $2p_{3/2}$, $1f_{5/2}$, $2p_{1/2}$ and $1g_{9/2}$ orbitals, respectively.
In this case, the protons were prevented from occupying the $1\pi g_{9/2}$ orbital
and eight neutrons were frozen in the $1\nu f_{7/2}$ orbital.
Electric quadrupole transition rates were calculated
using neutron and proton effective charges of $0.5e$ and $1.5e$, respectively.
Magnetic dipole transition rates were calculated
either with the bare neutron and proton orbital and spin $g$ factors
[$g_\ell(\nu)=0$, $g_\ell(\pi)=1$, $g_s(\nu)=-3.826$, $g_s(\pi)=5.586$, in nuclear magneton]
or with the spin values quenched by a factor 0.7.

The results of the calculations
are compared with the data in the middle and right panels of Fig.~\ref{f_eb}.
A few comments are in order.
It is clear that both interactions show effects of a shell closure at $N=40$.
This observation is trivial in the $pf$ model space
but even with the {\sl pfg} interaction
the excitation energies of the $3/2^-_1$ and $9/2^-_1$ levels are slightly higher
and the $B$(E2) values of the transitions to the ground state
are somewhat smaller in $^{67}$Co compared to lighter Co isotopes,
indicative of some effect of an $N=40$ subshell closure in the calculation.
Our measured $B({\rm E2};3/2^-_1\rightarrow7/2^-_1)$ value in $^{63}$Co
falls well below its value calculated with {\sl pfg},
which indicates that the shell model
overestimates the collectivity of the $3/2^-_1$ state.
On the other hand,
the calculated $B({\rm E2};9/2^-_1\rightarrow7/2^-_1)$ value in $^{63}$Co
comes out at 11.1~W.u,
in excellent agreement with the experimental value of 12.2~(54)~W.u,
obtained under the assumption of pure E2.
However, although the E2 rate is well reproduced by the shell model,
it fails to obtain the correct $B$(M1) value for this transition.
This can be inferred from the lifetime of the $9/2^-_1$ level
which is calculated to be 0.09~ps,
an order of magnitude shorter than the measured value.
This short theoretical lifetime
is due to a large calculated $B({\rm M1};9/2^-_1\rightarrow7/2^-_1)$ value of 0.36~W.u,
calculated with bare nucleon $g$ factors.
Quenching of the spin $g$ factors
does not alter this result significantly.
Because of the large uncertainty on our experimental value
for the lifetime of the $9/2^-_1$ level,
one might question the significance of this discrepancy.
We remark nevertheless that in the neighboring isotope $^{61}$Co
a similar situation occurs
since the measured lifetime of 0.8~(2)~ps for the $9/2^-_1$ level at 1285~keV~\cite{regan_yrast_1996}
is at variance with the calculated one of 0.08~ps---a
discrepancy again due to a large
$B({\rm M1};9/2^-_1\rightarrow7/2^-_1)$ value in the shell-model calculation.   

To what extent are the simple interpretations,
mentioned in the introduction,
borne out by the shell-model calculations?
More precisely, energy systematics would have us believe
that the $7/2^-_1$ and $9/2^-_1$ states
are a $(1\pi f_{7/2})^{-1}$ proton hole coupled to the $0^+_1$ or $2^+_1$ in Ni,
\begin{eqnarray}
|7/2^-_1({\rm Co})\rangle&\approx&
|(1\pi f_{7/2})^{-1}\times0^+_1({\rm Ni});7/2\rangle,
\nonumber\\
|9/2^-_1({\rm Co})\rangle&\approx&
|(1\pi f_{7/2})^{-1}\times2^+_1({\rm Ni});9/2\rangle,
\label{e_approx1}
\end{eqnarray}
while the $3/2^-_1$ state is rather a $1\pi f_{7/2}$ proton coupled to the $2^+_1$ in Fe,
\begin{equation}
|3/2^-_1({\rm Co})\rangle\approx
|1\pi f_{7/2}\times2^+_1({\rm Fe});3/2\rangle,
\label{e_approx2}
\end{equation}
where it is implicitly understood
that the neutron numbers on the left- and right-hand sides are the same.
If these approximations were true,
they would lead to simple selection rules for electromagnetic transitions.
For example, no M1 transition would be allowed
between the $9/2^-_1$ and $7/2^-_1$ states~(\ref{e_approx1})
since they involve different core states
that cannot be connected by the M1 operator
($0^+_1$ and $2^+_1$ in Ni, respectively).
These selection rules would hold,
irrespective of whether the states in the even-even cores
have a simple description in the shell model or not.
Analyzing the shell-model wave function of the $0^+_1$ state in $^{64}$Ni,
calculated with the {\sl gxpf1a} interaction,
one finds that the 8 protons are not confined to the $1\pi f_{7/2}$ orbital
but that they also occupy the $2\pi p_{3/2}$, $1\pi f_{5/2}$, and $2\pi p_{1/2}$ orbitals
with fractional occupancies of 7.50, 0.37, 0.10, and 0.03, respectively.
Similarly, the fractional occupancies for the $2^+_1$ state in $^{64}$Ni
are 7.38, 0.48, 0.11, and 0.03.
These states have an even more intricate structure
as regards the distribution of the neutrons over their available orbitals.
This complicated wave function of the even-even core states notwithstanding,
a structure for the $9/2^-_1$ and $7/2^-_1$ states of the form~(\ref{e_approx1})
would guarantee a vanishing $B$(M1) value between them.

Concerning the electromagnetic transition
between the $3/2^-_1$ and $7/2^-_1$ levels in $^{63}$Co,
we note that its measured $B$(E2) value,
well below the result obtained in both large-scale shell-model calculations,
is not inconsistent with the simple interpretations~(\ref{e_approx1}) and~(\ref{e_approx2}).
If the initial state in this transition
is a $1\pi f_{7/2}$ proton particle coupled to the $2^+_1$ in Fe
and the final state
is a $(1\pi f_{7/2})^{-1}$ proton hole coupled to the $0^+_1$ in Ni,
one expects in fact a reduction of E2 strength
as compared to what is observed for the $2^+_1\to 0^+_1$ transition in Fe.

In a shell-model calculation expansions of the type
\begin{equation}
|J_i\rangle=\sum_{jL_k}c^{J_i}_{jL_k}|j\times L_k;J\rangle,
\end{equation}
where $|J_i\rangle$ is a state in an odd-mass nucleus
formed by coupling a particle in orbital $j$ to even-even core states $|L_k\rangle$,
can be obtained by computing spectroscopic factors.
The coefficients $c^{J_i}_{jL_k}$ are found from
\begin{equation}
\left(c^{J_i}_{jL_k}\right)^2\approx
\frac{1}{\langle n_j\rangle}
\frac{\langle J_i||a^\dag_j||L_k\rangle^2}{2J+1},
\end{equation}
where the reduced matrix element of the creation operator in orbital $j$
and the occupancy $\langle n_j\rangle$ of this orbital for the state $|J_i\rangle$
are given by the shell-model code.
If we wish to express a shell-model state $|J_i\rangle$
as a hole in orbital $j$ coupled to an even-even core,
\begin{equation}
|J_i\rangle=\sum_{jL_k}c^{J_i}_{j^{-1}L_k}|j^{-1}\times L_k;J\rangle,
\end{equation}
the expression for the coefficients $c^{J_i}_{j^{-1}L_k}$ is
\begin{equation}
\left(c^{J_i}_{j^{-1}L_k}\right)^2\approx
\frac{1}{\langle\bar n_j\rangle}
\frac{\langle L_k||a^\dag_j||J_i\rangle^2}{2J+1},
\end{equation}
where $\langle\bar n_j\rangle$ is the `emptiness' of orbital $j$ for the state $|J_i\rangle$,
$\langle\bar n_j\rangle=2j+1-\langle n_j\rangle$.

Following the above procedure,
we find for the shell-model wave functions of states in $^{63}$Co,
calculated with the {\sl gxpf1a} interaction,
dominant components of the form
\begin{widetext}
\begin{eqnarray}
|7/2^-_1\rangle&\approx&
0.62|(1\pi f_{7/2})^{-1}\times0^+_1({\rm Ni});7/2\rangle
\pm0.38|(1\pi f_{7/2})^{-1}\times2^+_1({\rm Ni});7/2\rangle
\pm0.23|(1\pi f_{7/2})^{-1}\times2^+_2({\rm Ni});7/2\rangle+\cdots,
\nonumber\\
|9/2^-_1\rangle&\approx&
0.61|(1\pi f_{7/2})^{-1}\times2^+_1({\rm Ni});9/2\rangle
\pm0.31|(1\pi f_{7/2})^{-1}\times4^+_2({\rm Ni});9/2\rangle
\pm0.25|(1\pi f_{7/2})^{-1}\times2^+_2({\rm Ni});9/2\rangle+\cdots,
\nonumber\\
\label{e_decomp1}
\end{eqnarray}
and
\begin{equation}
|3/2^-_1\rangle\approx
0.55|1\pi p_{3/2}\times0^+_1({\rm Fe});3/2\rangle
\pm0.19|1\pi p_{3/2}\times0^+_2({\rm Fe});3/2\rangle
\pm0.18|1\pi p_{1/2}\times2^+_2({\rm Fe});3/2\rangle+\cdots.
\label{e_decomp2}
\end{equation}
\end{widetext}
Since the signs of the components are unknown,
they are indicated by $\pm$.
The components in the expansions~(\ref{e_decomp1})
represent barely 50 to 60\% of the total.
There is thus considerable fragmentation of the wave function
in the shell model
and, accordingly, the simple interpretations mentioned above
at best are only qualitatively valid in this model.
This is even more so for the expansion~(\ref{e_decomp2}) for the $3/2^-_1$ state
where the three largest components only carry 37\% of the total.
Consequently, the simple structure~(\ref{e_approx2})
is completely absent from the shell model
and on the basis of the expansion~(\ref{e_decomp2})
one would rather conclude that
a proton particle coupled to core states in Fe
is {\em not} the appropriate basis to interpret this state.

The large fragmentation also explains
the $B({\rm M1};9/2^-_1\rightarrow7/2^-_1)$ value
obtained in the shell model.
On the one hand, as explained above, no M1 transition is allowed between the main components
in the decomposition~(\ref{e_decomp1}).
On the other hand, the first component of the $9/2^-_1$ state
and the second component of the $7/2^-_1$ state
are connected by the proton part of the M1 operator.
Since the latter component is far from negligible,
this leads to the large calculated $B$(M1) value.
Given that the measured lifetime of the $9/2^-_1$ level
implies a $B({\rm M1};9/2^-_1\rightarrow7/2^-_1)$
which is a factor 10 smaller than the calculated value
(in both $^{61}$Co and $^{63}$Co),
a tentative conclusion of our analysis
is that the shell model predicts too much fragmentation.

\section{Conclusion}
\label{s_conc}
In summary, lifetimes of yrast states in $^{63}$Co and $^{65}$Co
were measured in a recoil distance Doppler shift experiment
and the reduced transition probabilities obtained from these lifetimes
were compared with large-scale shell-model calculations.
While the observed energy systematics suggest
that the $9/2^-_1$ state in $^{A}$Co can be interpreted
as a proton hole $(1\pi f_{7/2})^{-1}$ coupled to the $2^+_1$ state in $^{A+1}$Ni
and that the $3/2^-_1$ state can be interpreted
as a proton particle $1\pi f_{7/2}$ coupled to $2^+_1$ in $^{A-1}$Fe,
no such simple conclusions can be drawn on the basis of transition probabilities.
The $B({\rm E2};9/2^-_1\rightarrow7/2^-_1)$ value extracted for $^{63}$Co
is compatible with the $B({\rm E2};2^+_1\rightarrow0^+_1$) value in $^{64}$Ni
but the fact that the experimental value is obtained under the assumption of a pure E2 transition
renders any conclusion on this point provisional.
On the other hand, there is no apparent correlation
between the measured $B$(E2) values of the $3/2^-_1\rightarrow7/2^-_1$ transition in Co
and those of the $2^+_1\rightarrow0^+_1$ transition in Fe.
The fact that the former is consistently smaller than the latter
indicates that the $N=40$ gap has not as yet collapsed
in the Co isotopes we have investigated.
In addition, the analysis of the $3/2^-_1$ wave function
shows that this state as it is obtained in the shell model
cannot be written as a proton particle $1\pi f_{7/2}$ coupled to $2^+_1$ in Fe.
The shell-model calculations do reproduce many experimental results
but our analysis suggests that the {\sl gxpf1a} and {\sl fpg} interactions
predict too much fragmentation,
resulting in a significant overestimate of the $B({\rm M1};9/2^-_1\rightarrow7/2^-_1)$ value.
It will therefore be of interest to compare our experimental findings
with shell-model calculations with the new interaction
which recently has been devised for this mass region~\cite{lenzi_island_2010}.
Our results also call for the measurement of E2 transition probabilities in $^{65,67}$Co
which would clarify the evolution of the collectivity towards $^{68}$Ni.

\begin{acknowledgments}
We wish to thank F.~Nowacki for his help with the shell-model calculations.
This work was partially supported by the Agence Nationale de Recherche, France,
under contract nr ANR-07-BLAN-0256-03 and by the LEA-COPIGAL.
A.G. and G. de F. acknowledge the support on IN2P3, France, and MICINN, Spain, through the AIC10-D-000429 bilateral action. A.G. activity has been partially supported by the MICINN and Generalitat Valenciana, Spain, under grants FPA2008-06419 and POMETEO/2010/101.
\end{acknowledgments}

\bibliographystyle{apsrev4-1}
\bibliography{cobalt_resub.bib}

\end{document}